\documentclass[a4paper,11pt]{article}
\usepackage{generic,btx,times,url}

\usepackage{amssymb}
\usepackage{amsfonts}
\usepackage{latexsym}
\usepackage{epsfig}

\newlength{\expandwidth}
\newlength{\expandheight}
\title{Towards an Integrated Conceptual Modelling Kernel\\
       for \\
       Business Transaction Workflows{\footnote{Part 
of this work has been supported by CITEC, a business unit 
of the Queensland Government's Department of Public Works 
and Housing (formerly the Administrative Services
Department).}}}

\author{A.P. Barros${^1}$,
        A.H.M. ter Hofstede$^2$,
        H.A. Proper${^2}$ \\[0.3cm]
        \begin{tabular}[t]{c@{\extracolsep{2em}}c}
        \small $^1$School of Information Technology &
        \small $^2$Faculty of Information Technology \\
        \small The University of Queensland &
        \small Queensland University of Technology \\
        \small Brisbane, Qld 4072 &
        \small GPO Box 2434, Brisbane 4001 \\
        \small Australia &
        \small Australia \\
        \small & e-mail: E.Proper@acm.org
        \end{tabular}
        }
\date{}

\begin{document}

\maketitle

\begin{quote}
\begin{tabbing}
{\bf Keywords:} \= Conceptual Modelling, Business Process, Workflow, \\
                   Object-Oriented Analysis
\end{tabbing}
\end{quote}

{\sc Published as:}
\begin{quote}
  A.P. {Barros}, A.H.M.~ter {Hofstede}, and H.A.~(Erik) {Proper}. {Towards an Integrated Conceptual Modelling Kernel for Business Transaction Workflows}. Technical report, Department of Computer Science, University of Queensland, Brisbane, Queensland, Australia, November 1996.
\end{quote}

\begin{abstract}

\end{abstract}
\section{Introduction}
\label{intro}

The workflow concept, proliferated through the recently emergent
computer supported cooperative work (CSCW) systems and workflow
systems (see surveys in
\cite{Report:94:Fitzpatrick:CSCW,%
Book:93:Wastell:CSCW,Article:91:Rodden:CSCW}
and
\cite{Article:95:Georgakopoulos:WFOverview} respectively),
advances information systems (IS) implementation models by incorporating
aspects of collaboration and coordination in business processes. Under
traditional implementation models, applications are partitioned into
discrete units of functionality, with (typically) operational procedures
used to describe how human and computerised actions of business processes
combine to deliver business services. Through an endowment of business
process execution semantics, workflows permit a greater organisational
fit of ISs. Moreover workflows are specified at a level above traditional
applications, enabling program binding and access to a loosely-coupled set
of databases and files. Therefore, newer applications may be developed out
of existing applications to reflect reengineered business processes.

Crucial to the specification of any IS implementation is the \EM{conceptual}
level. This, of course, orients the analysis of a given domain towards its
essence (deep-structure) rather than to aspects of implementation
(physical-structure) or representation (surface-structure).
It is a well-known fact the later problems and
inadequacies are detected in specifications, the greater the expense of
correction
\cite{Book:90:Davis:SoftReq}.
For workflows, the standardisation of concepts is progressing
through the Workflow Management
Coalition{\footnote{Refer to http://www.aiai.ed.ac.uk/WfMC/index.html for
more details.}}.
While the set of terms and references defined so far characterise
sufficiently the notion of workflow, e.g.\ event, process (including
pre-conditions, post-conditions and state transitions) and organisational
(or actor) role, much of the focus is geared towards workflow management
systems and their specification languages. The emphasis is on part of
business processing, namely process execution semantics: sequence, repetition,
choice, parallelism and synchronisation. A sound conceptualisation requires
not only this but also that process semantics, e.g.\ the messaging, database
updates and retrievals involved, to be explicitly captured.

In general, for the conceptual level, \EM{techniques} are available
under different paradigms, for the modelling of different
aspects of a business domain (see e.g.\
\cite{Book:88:Olle:Methodologies}). When integrated into well-formed
methods, integrated IS specifications - result.
A number of paradigms may be discerned for workflow modelling:
process-centric, e.g.\
\cite{Article:91:Dur:TaskActor};
state-centric, e.g.\
\cite{Article:95:Antonellis:OOA};
and actor-centric, e.g.\
\cite{Article:94:Dietz:TransModelling}
(based on the speech-act theory synthesis of
\cite{Article:80:Flores:ComMod}).
Moreover the use of business (or enterprise)
models, e.g.\ as deployed in requirements engineering methods
\cite{Article:95:Berztiss:ReqEng,%
Article:95:Loucopoulos:ReqEng,%
Article:94:Anton:ReqEng},
in design methods
\cite{PhdThesis:93:Ramackers:BehaviourMod}
and in CAiSE tools e.g.\ AD/CYCLE
\cite{Article:90:Mercurio:CAiSE},
provides an \EM{organisational embedding} whereby a workflow model's
components may be backtracked to its real-world counterparts.

Although, the field of conceptual modelling has become fairly mature, the
application of techniques, has, by and large, followed the intuition of the
developers of models. This, of course, involves an informal to formal transition.
With workflow specifications, this transition is reduced, however a greater
alignment is required between the workflow modelling cognition and business
processing cognition. Beyond the qualification of fundamental modelling concepts
(e.g.\ process) with organisational attributes (e.g.\ business service), the business
processing semantics need to be infused into the semantics of a technique such
that a workflow may be expressed and communicated adequately using that technique.
In absence of a universal organisational theory, much uncertainty exists as to
how effective conceptual modelling techniques are for business workflows; whether,
given the diversity of business processing, any generable prescription of a 
business processing cognition is in fact possible or desirable.

In general, in one form or another, the requirements which lead to effective
conceptual modelling are that: technique should adhere strictly to the conceptual 
level; should provide a high degree of expressive power; should at the same time 
facilitate comprehensibility; and should be backed up by a solid formal 
foundation whereby both the syntax and semantics are clearly defined. 
Equally importantly, a technique should be suitable for its problem
domain, meaning that its concepts and features reflect closely those of
the problem domain.

The focus of the paper is on the extension of conceptual workflow modelling
techniques for business suitability.
Of course, to speak of a general business suitability is
vague since there are many types of organisations and many types of business 
processing
\cite{Book:85:Davis:MIS}.
Therefore, particular attention is drawn to a specific type of (operational)
business processing which exhibits precise execution paths. As examples, 
the processing of insurance claims, bank loans and land conveyancing, are 
mission-critical in nature and are rarely undertaken without strict operational 
procedure. 
Also, multiple interactions with clients and external organisations are
typically needed to fulfill service requests.

In
\cite{Report:96:Barros:BusWFPrinciples},
some ground work for developing a more precise notion of business suitability
has already been established. In particular, business suitability principles
were elicited from an assessment of both classical and business-oriented 
techniques. The \EM{\Embed\ Principle} describes how a model should be 
backtracked to organisational elements. The \EM{\Validate\ Principle} identifies 
the need for scenarios, and in particular, a business transaction, as distinct 
from business process and business service, for workflow cognition. The 
\EM{\InfoHiding\ Principle} requires that business processing undertaken for 
business service requests should be insulated from the requests, and in so 
doing, motivates the need for an explicit treatment of business services 
within conceptual modelling. The \EM{\Combine\ Principle} requires that all 
concepts involved in workflow model enactment be simultaneously present in the 
model. Simultaneously absent in the assessed techniques were the combination 
of structural \EM{and} behavioural aspects of workflows, human to computer 
interaction and temporal aspects. Finally, the \EM{\ErrorHandling\ Principle} 
identifies the need for operational error handling to be catered for at the 
conceptual level, thereby incorporating the recovery management focus of 
\EM{transactional} workflows into a general exception handling. In this paper, 
a number of essential modelling concepts and features for business 
transaction workflows are developed. 
The approach taken is to develop a a \EM{kernel} technique so that ...

The paper is organised as follows. 
In section~\ref{new-principles},
the problems and suitability principles are defined. In 
section~\ref{roadcl-btx}, useful concepts and features are combined, 
extended along 
the lines of the principles and applied to a real-world case study. In 
section~\ref{formal}, the kernel is described formally. Finally 
in section~\ref{conc}, the paper is concluded and further research issues
are identified.
\section{A business suitability synthesis}
\label{synthesis}

In this section, the main ideas behind \KerName\ are sketched using the 
business suitability principles: 
the \Embed\ (section~\ref{embed}); \Validate\ (section~\ref{validate});
\InfoHiding\ (section~\ref{infohiding}); \Combine\ (section~\ref{combine});
\ErrorHandling\ (section~\ref{errorhandling}).

\subsection{Organisational embedding}
\label{embed}

The \EM{\Embed\ Principle} requires that a technique ``embed all concepts 
in a conceptual model, directly or indirectly, but without redundancy, 
into organisational concepts''. This not only addresses the arbitrary 
relationship that can exist between conceptual models and their problem 
domains, but also the situation where specific business world views adopted 
in techniques, typically enterrpise models, prescribe the essential structure
of the conceptual model. The latter is, of course, a violation of the 
Conceptualisation Principle. Moreover, it precludes the design of an IS 
process from a network of business processes without first composing an 
``artificial'' business process. This restriction is evident in specification
frameworks of integrated modelling techniques, e.g.\
\cite{PhdThesis:93:Ramackers:BehaviourMod},
and CAiSE tools, e.g.\ AD/CYCLE
\cite{Article:90:Mercurio:CAiSE}.
Such a composition at the IS level and not the business level
relates to the comprehensibility of a process model, or
reflects IS design requirements (of a multi-organisational
domain say).

To facilitate organisational embedding, a business processing universe 
is defined which consists of a set of organisational concepts. 
(Although not described here) these include: organisational unit, actor and
actor roles, service, event, message, process and object type. For a 
particular area of interest (for IS development), three delineations are 
imposed over this universe. This is a business scope, a business domain 
and a business environment.
The business scope, is the broadest perspective possible for the area of 
interest, encompassing both the business domain and the business 
environment. 
The business domain is the primary area of interest requiring all the 
necessary concepts for a detailed description of business processing.
A business environment only requires those concepts which describe the
business processing interaction with the business domain. For example, 
the events and their messages, are sufficient for understanding interactions 
to and from a business domain.
Through a consistent perception of a business processing universe, its 
detailed projection through a business domain and its partial projection 
through a business environment, a ``way of thinking'' is also apparent. That 
is, for an area of interest, the separation of the clients (environment), the 
server (domain) and the ancillary servers (environment) at the outset,
clarifies the modelling detail required. 

\subsection{Scenario validation}
\label{validate}

The \EM{\Validate\ Principle} requires that ``a technique should provide 
an explicit notion of scenario for model validation''. Validation is 
concerned with the interpretation of a conceptual model's domain semantics
as distinct from verification which is concerned with formal syntax and
semantics. 
In a general sense, a scenario represents an \EM{effect} in a business 
domain for some well defined reason, which is expressible using the 
concepts (of typically more than one partial model). 
In business processing terms, it is clear that a reason may be associated
with an event, and the effect is the resultant triggering of a set
processes through which information may be accessed. Furthermore events, 
possibly invoking processing in the business environment may also result. 
Ultimately, the occurrence of some final event signifies the (logical) 
termination of processing; i.e.\ representing the organisation's
recognition that no further processing should proceed.

While scenarios appear synonymous with workflows, there is a subtle
distinction. A scenario is a concept which serves interpretation, while
a workflow is a concept which  serves implementation. 
Depending on the degree of complexity of a workflow, a scenario may
address part, or all of a workflow, and possibly several workflows. In
accordance with the \EM{\Embed\ Principle}, the notion of a
\EM{business transaction}, drawn from a Macroeconomics perspective of 
organisations 
\cite{Book:91:Boyes:BusMod},
is proposed as a scenario.
Business transactions are accountable units by which an organisation
measures its exchange of (goods and) services. Ultimately, this measure
is aggregated for a nation's economic metrics {\footnote{Gross 
National Product (GNP) and Gross Domestic Product (GDP).}}.
As a subunit of business processing, indeed one which represents a
cost closure, a business transaction provides a well-recognised mechanism
for workflow model interpretation. The term business transaction 
will hereafter be preferred over business processing, and the notion
\EM{business transaction workflow} will qualify that class of
business processing being considered for workflow implementation.

\subsection{Service information hiding}
\label{infohiding}
A key distinction in business transactions is that between business
services and business processes. This is, in fact, a generalisation
of the distinction between events and processes. An event, afterall,
is associated with some intention, and more than one event may share
the same intention. In a business sense, intentions are denoted by 
business services. 

As such, business services are a described (external) organisation 
of functionality which do not do anything as such, other than being 
associated with process interactions. This may involve external access 
such as client requests and responses from outside organisations, or 
internal access to services in different parts of an organisation. 
Inherently, they have a set of \EM{states}, e.g. initiated,
processing, rejected, and each state is associated with, and dependent on,
a particular course of action resulting from a particular event. 
Business processes on the other hand, are a 
prescribed (internal) organisation of functionality reflecting the 
mechanisms by which business services are delivered. Unlike services, 
they perform concrete actions, (e.g. data transformations, updates and 
retrievals) and their states are (relatively speaking) dependent on the 
success of their processing. 

The \EM{\InfoHiding\ Principle} requires that ``a technique should allow 
the formulation of service requests to be independent of their actual 
processing''. This is to avoid the problem of the \EM{direct} triggering
of processes given the context of triggering. From the point of
view of the environment or from different parts of an organisation, the
actual business processes triggered for some business service request are 
inconsequential for the formulation of the request. That is, the request is 
issued for a business service and as a result some internal mechanism is 
used to determine what action to take. An implication is that when 
processes are reengineered, the actual request is not affected. This, of 
course, is an application of the well-known \EM{Information Hiding Principle}
in software design. 

\subsection{Cognitive sufficiency}
\label{combine}
The \EM{\Combine\ Principle} relates the inclusion of all the concepts
which ``provide a sufficient cognition of a model such that no assumptions
about fundamental aspects of business processing \EM{execution} semantics 
are required''. Notable areas of variance in process and workflow
modelling techniques are addressed below.

\subsection*{Object and control flows}
Although techniques allow either structural or behavioural aspects of 
processes to be modelled (given their distinctive purposes of analysis), both 
are required for capturing a business transaction's execution semantics.
Structured process models, e.g.\ a data flow diagram (DFD)
\cite{Book:89:Yourdon:Analysis},
describe object (data) flows, i.e.\ identifiable \EM{containers} of objects, 
object stores, i.e.\ persistent repositories of objects, and object 
transforming processes. 
Behavioural process models, e.g.\ Task Structures
\cite{Report:91:Hofstede:TaskAlg},
describe process sequences, repetition, choice, parallelism and 
synchronisation.

Clearly, behavioural aspects are crucial for the specification of execution 
semantics. Structural concepts are also important for two reasons. Firstly, 
object flows are transmitted by triggers related process interactions, 
i.e.\ processes and processes, and events and processes. As containers of 
objects, they should be differentiated from the set of attributes they 
transmit. 
For example, a process precondition should make reference to a particular 
object flow type, when other object flow types have the same set of attribute 
types as it. 
Secondly, process specifications would be incomplete without the incorporation 
of the object transformations and objects stores involved.

\subsection*{HCI}
Another area of cognitive insufficiency is the conceptualisation of 
human to computer interaction (HCI) by techniques. Although traditionally
determined at a detailed design stage, HCI \EM{points} and their 
\EM{dialogues} may be defined in a process model to reduce the 
\EM{waterfall} between conceptual, design and implementation levels  
\cite{Report:89:Brinkkemper:DialogueSpec}. 
Moreover, it is possible to derive them from a business transaction's
semantics; e.g.\ in
\cite{PhdThesis:93:Ramackers:BehaviourMod},
if a human and computerised actor types involved in an action, that action
is refined a method of an external object type (a form). (How do our
business transaction semantics extend this idea?).

\subsection*{Temporal aspects}
Despite the varying support in techniques, business transactions require
temporal specifications for process dependencies, and therefore process  
pre- and post-conditions. Direct analogues may be found with normal
process execution: a process may be required to execute within a time
duration of another process (sequence); a process may execute
repeatedly within a time duration (repetition); a process may be required
to execute either at one time or another (choice); a number of processes
may be required to execute simultaneously, at some time (parallelism);
or messages from a number of processes may be required within a given
time (synchronisation). 

\subsection{Execution resilience}
\label{errorhandling}

A more specialised aspect of execution semantics relates to \EM{execution
resilience}. That is, errors can occur which affect the normal execution
of a business transaction. Of course, error prevention may be defined 
through database constraints and process pre- and post-conditions while 
model verification eliminates erroneous specifications. 
However, \EM{operational} errors can still occur beyond the control 
of an IS. For example, clients may not abide by business processing
rules (borrowers not returning items by their due-dates to a library), 
or system crashes, may occur. 

The \EM{\ErrorHandling\ Principle} requires that ``a technique should 
support the handling of operational errors, so that business processing 
execution may be verified as being resilient".
Although traditional process modelling techniques do not deal with this 
aspect, a recently proposed workflow modelling technique, e.g.\  
\cite{Article:95:Casati:ConceptualWF},
provides basic mechanisms for exception handling. A more detailed treatment
of  non-deterministic failures has recently become the subject of workflow 
implementation specifications. In particular, the database transaction model 
(more recently described in 
\cite{Book:93:Gray:TransMgmt})
with its ACID properties (atomicity, consistency, 
isolation and consistency) has been extended for workflow execution 
semantics; see survey of transactional workflows issues in
\cite{Book:94:Kim:TransMgmt} (pp. 596-598). 
Under it, a transaction binds a set of database 
operations into an atomic unit of execution. Following the requirement 
of \EM{failure atomicity}, a transaction's changes to a database(s) are 
\EM{committed} if the execution is successful or \EM{rolled-back} if not. 
Following the requirement of \EM{execution-atomicity}, the concurrent 
execution of transactions should have the same effect as if they were 
executed in a \EM{serialisable} order.
Workflows are more complex structures than database transactions, and 
it is unacceptable that the failure of any one of its processes results 
in the rollback of the entire workflow. 

Like traditional transactions, the notion of \EM{commit} points 
can be used to define the atomic unit of workflow execution and 
recovery. The normal place for the occurrence of a commit would be 
expected to be at the end of a process's execution. However, the atomic 
unit can be extended to include more than one process. Of course, process 
objects can be composite in which case the entire superprocess can be 
regarded as an atomic unit or atomic units can be formed out its 
decomposed processes.

The broader question of the appropriateness of database transaction 
rollback for workflow recovery then emerges. For one, workflow 
processes are more sophisticated than database transactions, 
e.g.\ they may involve messaging or they may be human-oriented tasks. 
Certainly, for a conceptual level specification, it seems to make more 
sense to describe what the desired action should be rather than 
having to include its implementation strategy.
Clearly, during a crash type of scenario, the desired action is a
\EM{redo} of the crashed process. A redos only strategy, as applied 
in database management system (DBMS) implementations, is 
associated with a \EM{rollforward} (i.e.\ a forwards) recovery, aimed at 
database transaction durability. We therefore propose a rollforward 
recovery strategy for failures situations. A special type of abort 
message is dedicated for this, i.e.\ a failure abort.
Additionally, like for distributed transaction management, we
recognise the need for \EM{contingencies}. Contengencies are
run when transactions fail to start, for whatever reason. Similarly,
we adopt contingent processes in workflows with a basic
extension which allows a range of contingencies for given
numbers of failure. This enhances the robustness of contingency, and
allows for the \EM{forcibility} of a process (i.e.\ having an. 
As in distributed transaction management, this means that the process 
should succeed eventually.

We contend that a \EM{rollforward} recovery strategy be adopted
when workflows are aborted as a result of application-generated
errors. We have introduced another type of abort 
message to convey this, i.e.\ a non-failure abort. Following
advanced transactions models 
\cite{Article:90:Agrawal:DBTrans}
which allow \EM{nested} transactions, we adopt \EM{compensations}.
Compensations are an ``undo'' mechanism since
subtransactions commit and release their 
resources prior to the parent transaction reaching a commit state.
If the parent transaction is aborted, unwanted committed data will
exist in a database. We appropriate such a strategy for workflow rollback 
recovery since the rollback can be applied over committed processes.
Examples of compensations in a workflow could include: the logical-reverse 
of an update, i.e.\ an idempotent update; sending abort messages to remote 
services for notification; triggering transitions to erroroneous service 
states.

\section{Business transaction modelling: a practical perspective}
\label{roadcl-btx}

Having described its main ideas, the detailed modelling concepts and features
of \KerName\ are illustrated through examples of a real-world business 
transaction example. This is based on \EM{road closures} undertaken by the 
Queensland State Government's Department of Natural Resources. The full 
case study is described in
\cite{Report:96:Barros:WFCaseStudy}.
This case study provides the complementary empirical insight into
An overview of road closures is described in section~\ref{road-closure} while 
the business scope is described in section~\ref{scope}.
The application of \KerName's three fundamental models, the \EM{object}, 
\EM{process} and \EM{service} models, are described in section~\ref{object}.
section~\ref{process} and section~\ref{service} respectively.

\subsection{Background}
\label{background}

As a consequence of the Westminster System used in Australia, the government 
administration of land falls under a number of statutes (or legislative acts) 
which involve a number of statutory authorities. In Queensland, the State 
Government's Department of Natural Resources under the 
Lands Act{\footnote{Lands Act 1994; i.e.\ last issued in 1994.}} is 
commissioned to grant \EM{tenure} for unallocated state land and reserved 
land. In a broad sense, this includes: the granting of ownership through 
freehold titles; the granting of custodianship for some purpose through 
leasehold titles - leases; the establishment of reserves for national parks 
and wildlife etc.; and the dedication of roads (which by definition in the 
Land Act implies public use). These apply to \EM{parcels} of land which 
are composed of one or more elementary allotments, or \EM{lots}.

In order to grant tenure, the Department of Natural Resources obtains the 
views of the relevant stakeholders - statutory authorities, affected
landholders and affected associations in the community -  in order to 
determine whether proposed and potential use of the parcel affects the 
surrounding area's current and future land use and the current 
legislation. The statutory authorities include local governments, 
electricity power suppliers, telecommunications carriers and environment 
and heritage regulation authorities.

The process of determining whether tenure should be granted is complex
taking from days up to months. During this period, repeated checks are 
required to ensure that the appropriate requirements are satisfied. These 
are needed since different actions constantly occur on related aspects of 
land. For example, during the process of investigating whether a mining 
lease should be granted, an overlapping part of the land may become 
heritage-protected while another overlapping part may be needed for a 
railway corridor. Clearly, the three tenures may result in incompatible land 
use. Furthermore, repeated interaction with the different stakeholders may 
be necessary to resolve unsatisfied requirements. During this period also, 
business processes may change due to changes in legislation as well as 
in organisational restructures. In short, an effective coordination of 
processing requires: the integrity of tenure grants to be preserved;
insulation from business process change; a minimisation of customer 
interaction.

\begin{figure}[htb]
\centering
\epsfig{figure=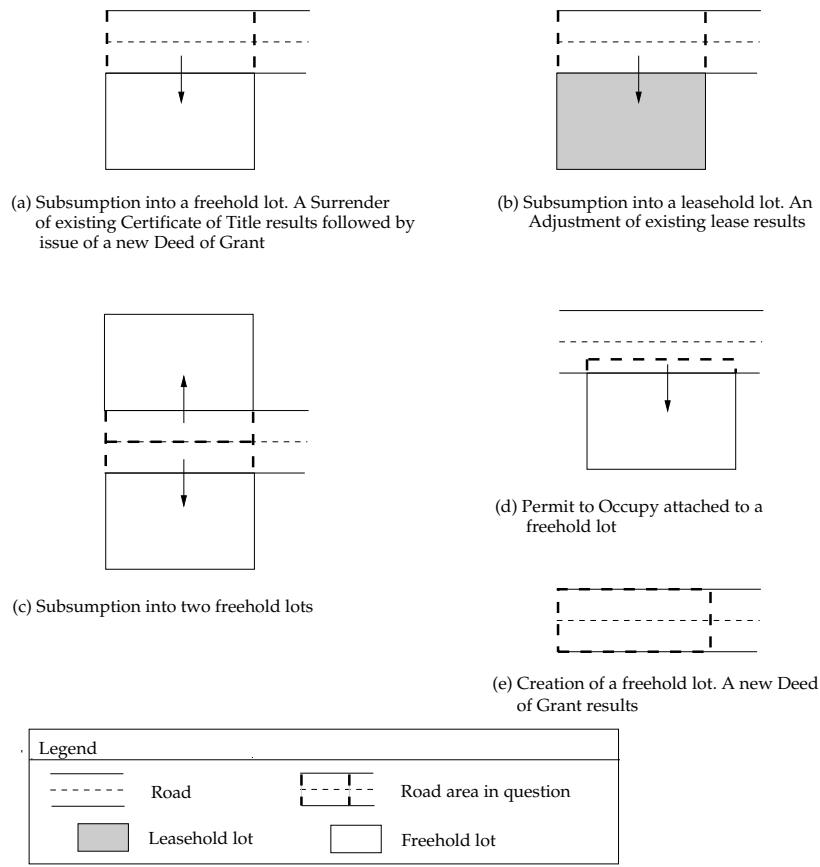,width=0.7\linewidth}
\caption{Road closure examples}
\label{roadcl.egs.fig}
\end{figure}

The closure of roads is a particular instance of tenure allocation. 
Under the Lands Act, the Minister for Lands approves road closures. This 
responsibility may be delegated to specific persons. Apart from the legal 
reasons, the approval (and for that matter the disapproval) of road closures 
carries important ramifications. For one, the general public have certain 
rights and expectations to roads. A road should not be closed for 
reasons which include the current or potential blockage of dedicated access 
to another parcel(s), the compromise of the transportation network, 
environmental degradation (e.g. some roads form important corridors and 
refuges for flora and fauna) and the existence of Native Title{\footnote{Under 
the Native Titles Act (Queensland) 1993, ammended in 1994, parcels without 
any allocated tenure are deemed to have Native Title, 
i.e.\ their use is determined by the Aboriginal people of Australia.}}.
Figure~\ref{roadcl.egs.fig} contains different examples of road closures.

A road may be closed permanently or temporarily. If permanent, it may be 
subsumed into one or more adjoining parcels as depicted in 
\textsf{(a)}, \textsf{(b)} and \textsf{(c)}. 
A subsumption into a freehold tenured parcel \textsf{(a)} involves a 
\EM{Surrender} of the existing title, i.e.\ a \EM{Certificate of Title}, and 
the issuing of a new title, i.e.\ a \EM{Deed of Grant}{\footnote{A Deed of Grant 
is a title which signifies the first tenure of a parcel.}} over the
``new'' parcel. 
A subsumption into a leasehold tenured parcel \textsf{(b)} involves 
an \EM{Adjustment} of the existing lease. For temporary closures, a 
\EM{Road License} (not illustrated) or a \EM{Permit to Occupy} 
\textsf{(d)}, may be issued over an area, where the State retains the right 
to re-dedicate the area at any subsequent time. In the case of a Permit 
to Occupy, the road does not lose its status and the public's access cannot 
be impeded completely, e.g. a side walk cafe. Roads may be created into 
parcels without subsumption \textsf{(e)}, in which case a Deed of Grant is 
issued. A complex road closure example (not illustrated) occurs during the 
development of new estates where a developer wishes to restructure a 
grid-like parcel arrangement into one which has a more irregular arrangement, 
typically with cul-de-sacs. This attracts more buyers. For this, the existing 
roads first need to be closed and the surrendered land needs to subsumed into 
the existing parcels. This is provided through the Road Closures business 
transaction. The newly designed road then needs to be opened, through another
business transaction. 

\subsection{Object Modelling}
\label{object}

The object (data) modelling technique of \KerName\ is based on the
Conceptual Data Modelling Kernel (CDM) \cite{Report:95:Proper:CDMKernel}.
It extends the Object-Role Model (ORM) kernel
\cite{Report:94:Brouwer:ORMKernel}
to not only allow modelling of ORM schemas, Entity-Relationship and
(structural) Object-Oriented schemas.
Its applicability is therefore fairly wide.
It achieves this through the provision of a generic mechanism for
\EM{abstraction}. In particular, set of object types can be clustered 
together to form abstractions for different structural constructs, 
e.g.\ specialisation heirarchies and aggregation. 

\begin{figure}[htb]
\centering
\epsfig{figure=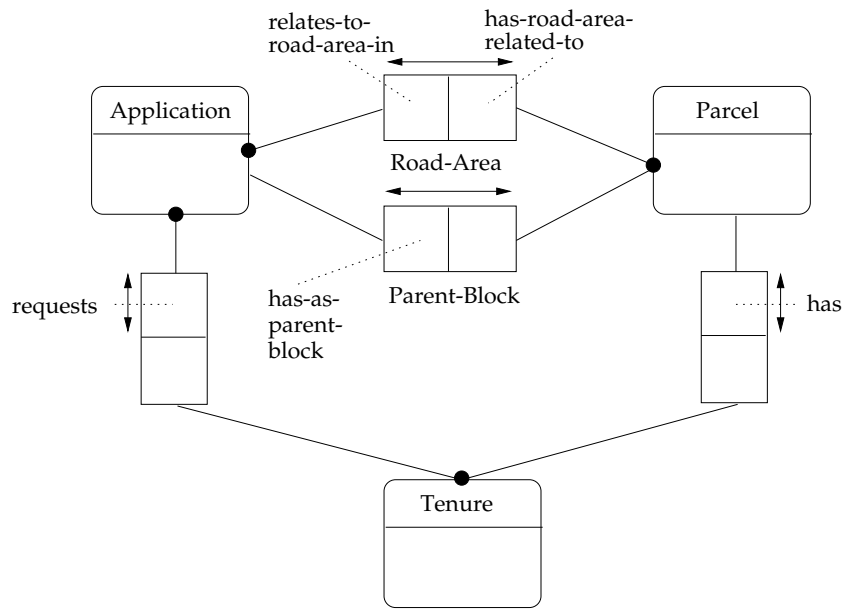,width=0.7\linewidth}
\caption{A highest level object model}
\label{topcdm.fig}
\end{figure}

The highest level object model for the road closures business 
domain is depicted in Figure~\ref{topcdm.fig}. The three object types,
\textsf{Application}, \textsf{Parcel} and \textsf{Tenure}, are of 
schema abstractions. These further decompose into object models. 

An \textsf{Application} plays a role, \textsf{relates-to-road-area-in}, with 
a \textsf{Parcel} through a binary relationship (a split rectangle with each 
box representing a role). The converse role \textsf{has-road-area-related-to}
is also depicted. A uniqueness constraint (a double headed arrow over 
both roles) indicates that this relationship is many-to-many while a mandatory 
role (a black dot) indicates that the corresponding role must be played. In 
other words, each \textsf{Application} must have at least one road area. 
A parent block may also pertain to an \textsf{Application}. Road areas and
parent blocks have the same classification, namely \textsf{Parcel}. The
junction of roles on the mandatory role constraint of \textsf{Parcel} 
indicates that it inclusively plays one role or the other. 
\textsf{Road Area} and \textsf{Parent Block} denote fact types; i.e.\
the object types and the relationship types involved. This allows fact
instances as a whole to be referred to. Fact type names will not be further
depicted. A \textsf{Parcel} (typically) \textsf{has} 
\textsf{Tenure} while an \textsf{Application} \textsf{requests} a 
\textsf{Tenure}. Both relationships involve a one-to-many 
uniqueness constraint (a double headed arrow over one role only).

\begin{figure}[htb]
\centering
\epsfig{figure=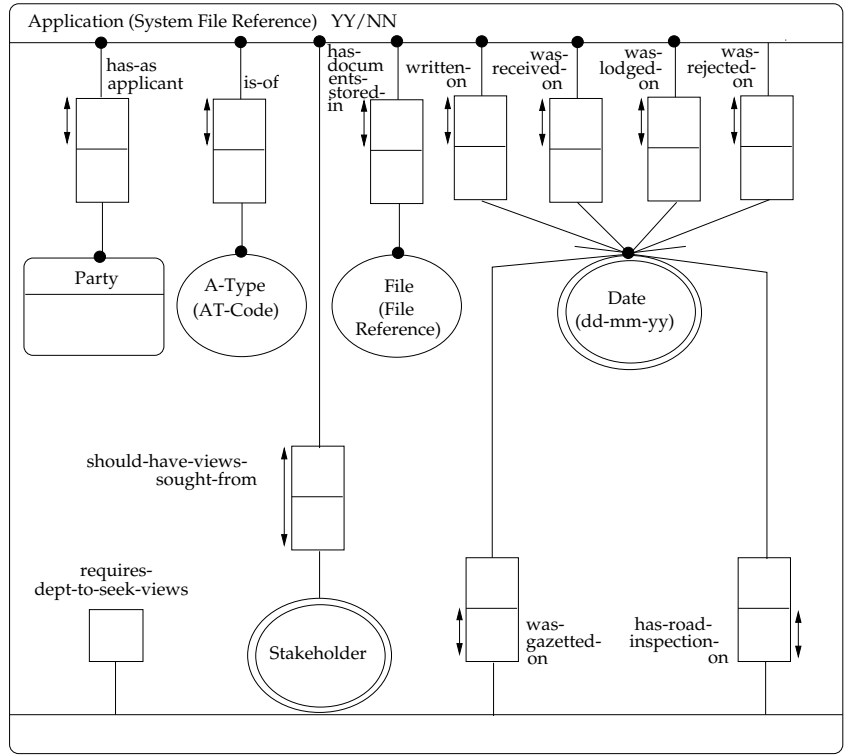,width=0.7\linewidth}
\caption{An object type decomposition}
\label{app.cdm.fig}
\end{figure}

None of the decompositions other than that for \SF{Application} in 
Figure~\ref{app.cdm.fig} are shown. At the top, the reference type is indicated 
in parentheses, namely \textsf{System File Reference}. It represents the way 
in which each \textsf{Application} is identified. \textsf{YY/NN} is the 
format for the reference, in this case, denoting a year followed a two digit 
number. Each \textsf{Application has-as-applicant} a \textsf{Party}
(i.e.\ a generic for a person or an organisation).
An \textsf{Application} also \textsf{is-of} a type, \textsf{A-Type}, 
\textsf{should-have-views-sought-from} a set of \textsf{Stakeholders} 
(a double lined object type indicates its occurrence elsewhere), 
\textsf{has-documents-stored-in} a \textsf{File}, has dates when it 
was \textsf{written-on}, \textsf{received-on}, \textsf{lodged-on}, 
\textsf{rejected-on}, \textsf{gazetted-on} and 
\textsf{has-road-inspected-on}, and an indication of whether the department 
is to seek views from the \textsf{Stakeholders}, described through the 
unary role \textsf{dept-to-seek-views}. The lines without roles attached to
\textsf{Date's} mandatory role constraint indicates that all roles (including
those not relevant in this schema) played involve mandatory inclusive 
disjunction.

\subsection{Process Modelling}
\label{process}

The \KerName\ process modelling technique extends Hydra's transaction
modelling
\cite{PhdThesis:92:Hofstede:DataMod}
which incorporates Task Structures and LISA-D, to include structured process
modelling (e.g.\ DFD) concepts, messaging similar to
\cite{Report:96:Hubbers:BehaviourSemantics}, 
complex decisions, temporal aspects and operational error handling.

\begin{figure}[htb]
\centering
\epsfig{figure=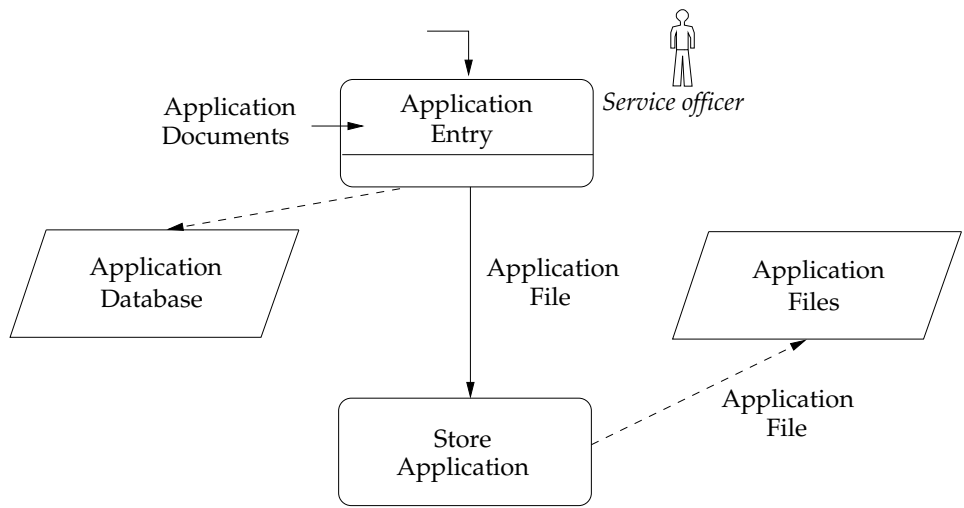,width=0.8\linewidth}
\caption{An example of structural processing extensions for a behavioural process model}
\label{app.lodge.fig}
\end{figure}

The processing of road closures commences with the arrival of a
letter of application and relevant documents, i.e.\
\SF{Application Documents}, at a \SF{Service Centre}. The
\SF{Application Lodgement} depicted in 
Figure~\ref{app.lodge.fig}, is the first of the three high-level business
processes to be executed. In it, an 
\textsf{Application File}, formed to contain the documents, is sent to 
the relevant \textsf{Regional Office} where it is filed away in the 
\textsf{Application Files} by \textsf{Store Application}. An actor role,
\textsf{Service officer}, is indicated for the \textsf{Application Entry}.
The convention we have adopted is that an actor role applies to the indicated
process object (a process or a decision) and the process objects which are
triggered subsequently, until another actor role is indicated.

The process model resembles, in part, a Hydra task Structure.
Processes (boxes) including an initially executed process (denoted 
by the bent arrow), together with execution triggering (arrows between 
processes) are shown. Through the depictions of \SF{Application
Documents} and \SF{Application File}, it can be seen that we have 
introduced messages. Both are hard-copy messages. 
\SF{Application Documents} has been depicted to arrive 
asynchronously (a small arrow embedded in a process). This means the 
the receipt (sending) of the message does not suspend any processing. 
(In section~\ref{service}, we will describe
the orthogonal issue of how the arrival of a message - as an event -
triggers the execution of a process). The messaging mode of
\SF{Application Files} from \SF{Application Entry} to \SF{Store
Application} is also asynchronous. We have adopted this variation of
notation for intra-model messaging between processes.

\begin{figure}[htb]
\centering
\epsfig{figure=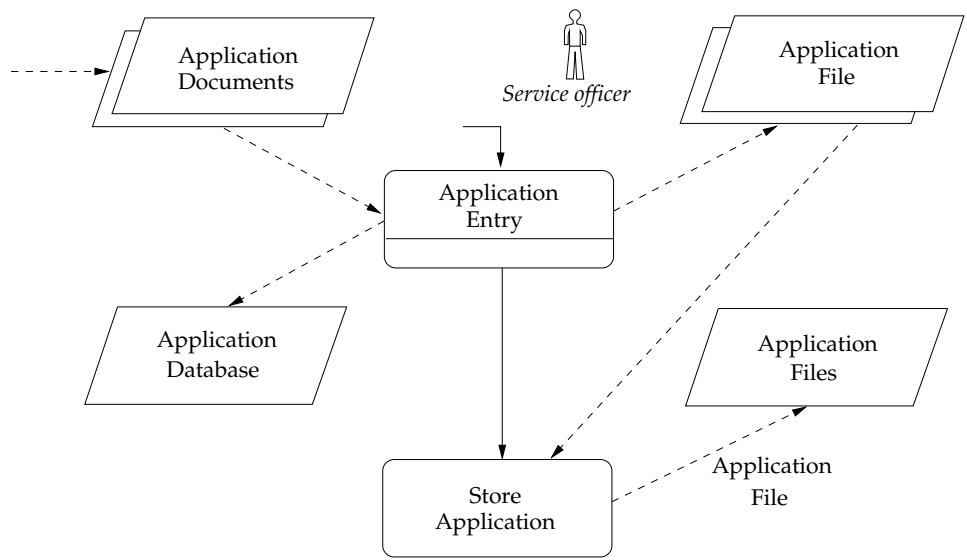,width=0.8\linewidth}
\caption{An example of explicit message buffering}
\label{alt.app.lodge.fig}
\end{figure}

In general, the semantics of messaging do not follow those of
triggering. In business transaction processing, afterall, messages are
not ``aimed'' directly at processes but at containers, e.g.\ an in-tray 
or a mail-box. The retrieval (sending) of messages from (to) containers 
is described in the process specification. To provide a treatement of 
\EM{transient} storage , we have extended the interaction of Hydra buffers. 
In Hydra, buffers permit a FIFO (first-in first-out) queuing protocol. In
\KerName, this has been generalised for any such protocol: LIFO (last-in
first-out), random or any predicate specifiable order.
An equivalent representation for Figure~\ref{app.lodge.fig} which makes 
message buffering explicit is depicted in Figure~\ref{alt.app.lodge.fig}. 
This representation is more preferable when the 
buffers add to the comprehensive value to the model. Notice, the name of a message 
need not be repeated throughout a process model. 
Naming on its initial occurrences (through messaging or creation in a process) 
and final occurrences (through messaging or storage) is sufficient. 
In this case,
the initial occurrence of \textsf{Application File} is created by a process
while the final occurrence involves storage. \textsf{Application
Documents} also occurs through inbound messaging but its final occurrence 
is not apparent, since it is contained in \textsf{Application Files}.

Included in process specifications are pre- and 
post-conditions and component actions for database access. 
A major benefit of using \EM{tightly-coupled}
integrated techniques such as Hydra lies in the increased expressiveness
available for their conceptual specification languages.
In general, such expressiveness is necessary to capture more fully
process semantics. As a basic example, consider the following LISA-D
formulation for capturing details of an \textsf{Application}: 

\[ \arraycolsep = 0pt
   \begin{array}{ll}
   & \RIDLADD \RuleVar{ Application:} \mbox{ Current-App} \RuleVar{ has 
System File Reference:} \mbox{ Current-File-Ref} \\
   & \RIDLADD \RuleVar{ Application:} \mbox{ Current-App} \RuleVar{ has-as-applicant Party} 
     \RIDLWITH \RuleVar{ Name} = \mbox{Current-Applicant} \\
   & \RuleVar{. . .} \\
   & \RIDLADD \RuleVar{ Application:} \mbox{ Current-App} \RuleVar{ lodged-on Date: \$Today}
   \end{array} \]

The schema assigment of \SF{Application} to the 
\SF{Application Database} makes this possible.
Of course, the variables Current-App, Current-File-Ref etc., 
permit temporary storage during the data entry. For the data entry,
HCI points have been introduced (slots on the lower part of process symbols). 
The description of the form/screen of the application being used, may be
contained in this slot.
The (actual) names have been omitted from the discussion (much like a 
CAiSE tool where certain textual details in a model may be ``clicked'' on 
and off). Of course, object schemas are required to be assigned to HCI 
points and messages. Like databases, they are storage entities and so
require an information grammar. 

\begin{figure}[htb]
\centering
\epsfig{figure=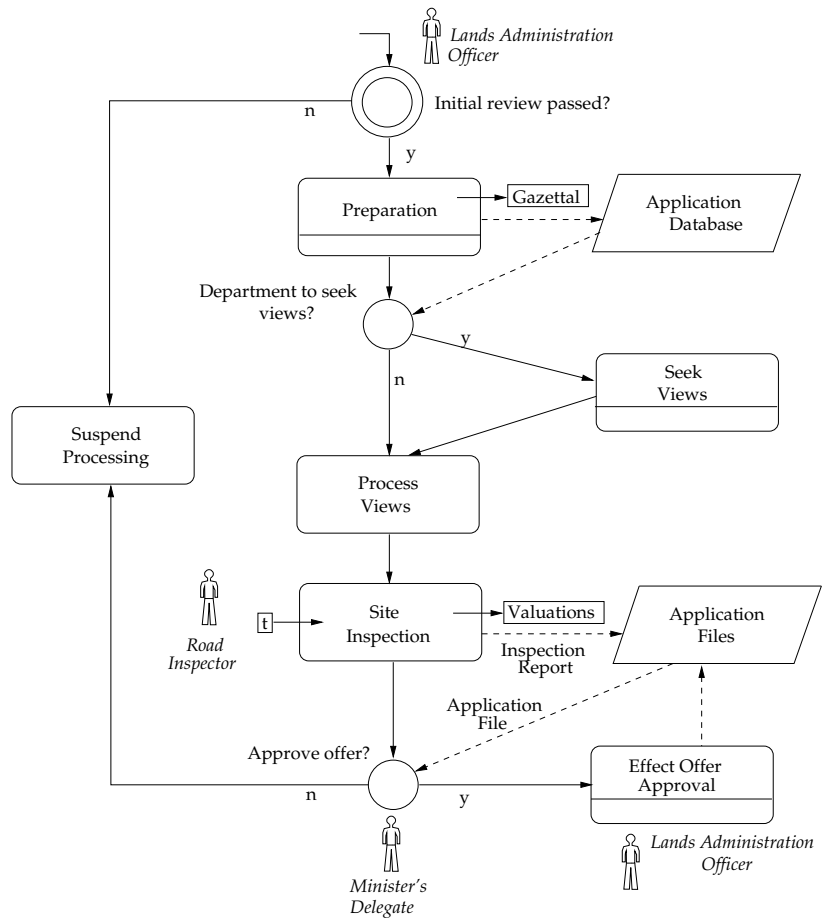,width=0.7\linewidth}
\caption{An example of further process modelling extensions}
\label{app.invest.fig}
\end{figure}

A more complicated process model is depicted in Figure~\ref{app.invest.fig}
for \SF{Application Investigation}. In brief, it
consists of a number of internal checks to determine whether the 
\textsf{Application} is valid, a \textsf{Preparation} for a more detailed 
investigation, \textsf{Seek Views} and \textsf{Process Views} of the 
\textsf{Stakeholders} as a part of the detailed investigation, 
and a \textsf{Site Inspection} as another part. A decision is then made to
\textsf{Approve (an) offer} which if negative, results in either a rejection or 
request for further information/action through \textsf{Suspend Processing}, or
if positive results in a preparation to make the offer through
\textsf{Effect Offer Approval}. In the description that follows, only
\SF{Initial Review Passed?}, \SF{Preparation} and \SF{Seek Views} are
further elaborated on.

\begin{figure}[htb]
\centering
\epsfig{figure=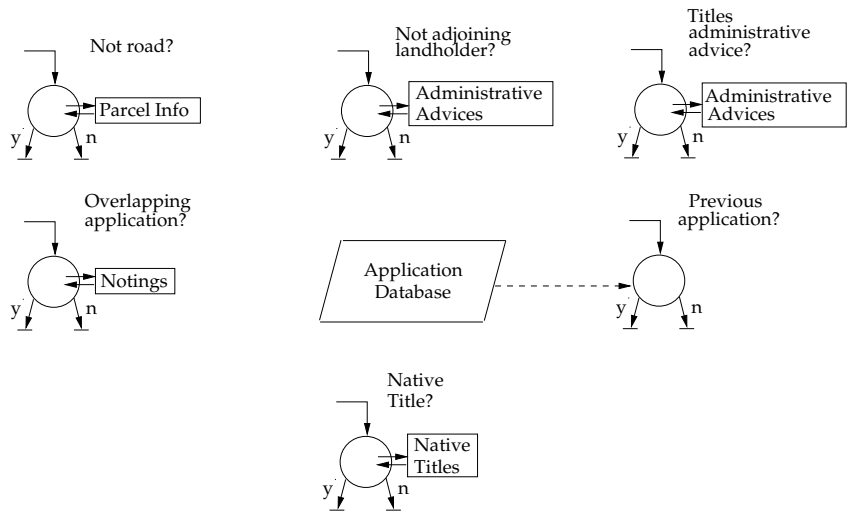,width=0.7\linewidth}
\caption{An example of a complex decision}
\label{review.fig}
\end{figure}

The example of the internal checks presents the need for an extension
to decision handling in traditional modelling. Under Hydra Task 
Structures for example, decisions yield either a positive or negative 
outcome, given their rules.
Moreover an outcome can terminate execution, returning control to
the supertask. It is evident through this part of road closures, as
depicted in Figure~\ref{review.fig}, that decisions in real-world business 
transactions may be based on sub-decisions. In this example, the \EM{complex}
decision is refined into \EM{simple} decisions, each of which is executed 
in parallel, with an implied synchronisation of their outcomes. 

\begin{figure}[htb]
\centering
\epsfig{figure=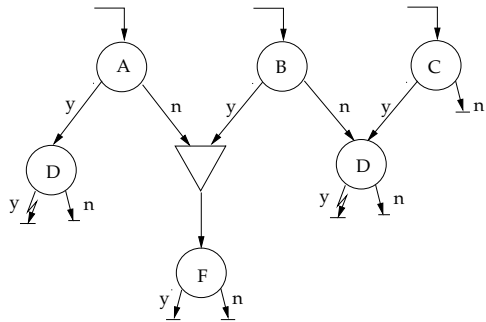,width=0.4\linewidth}
\caption{An example of a decision network based on a complex decision}
\label{compdecis.fig}
\end{figure}

It is possible to build up powerful complex decisions as depicted
in Figure~\ref{compdecis.fig}. This is an example of a decision
network consisting of simple and possibly further complex 
decisions with execution triggers (i.e.\ dependencies) between them. 
Unlike the example of Figure~\ref{app.invest.fig}, not all decisions 
need be evaluated as also seems evident in decision-making of
real-world business transactions, among others. For this, we propose
terminating \EM{aborts} for decision outcomes which when executed
terminate complex decision processing. Decisions \SF{D} and \SF{E}
have these. Furthermore, note the differences in decision dependency.
\SF{D} is triggered by an outcome of \SF{A}, while \SF{E} is
triggered by outcomes of \SF{B} or \SF{C} (and therefore may be invoked 
twice). On the other hand, outcomes from both \SF{A} and \SF{B} are required
for \SF{F}. The Task Structures synchroniser construct (triangle) 
caters for this.

A further extension to decision processing is the accommodation of 
messaging. In Figure~\ref{review.fig}, most decisions require data from
messages for the decision rules. Also the messaging of ``remote'' 
services (boxes attached to the messaging arrows) is illustrated.
Unlike the previously discussed form of messaging
which was \EM{asynchronous}, the depicted messages are \EM{synchronous}. 
That is, a message is sent out and an incoming message is anticipated 
(hence two embedded arrows). From the time that the message is sent out 
to the time that the message is received, no execution proceeds. Again
through LISA-D, we illustrate how highly expressive conceptual specification 
languages allow quite sophisticated
decision rules to be formulated. 
In general, LISA-D expressions are built from \EM{paths} between object 
type names and role names. 
The following decision rule pertains to the positive 
outcome of \SF{Previous Application?} (assuming a two year threshold):

\[ \arraycolsep = 0pt
  \begin{array}{ll}
  \RIDLLET \mbox{ previous-app } \RIDLBE \RuleVar{ Application(} & 
\RuleVar{has-as-parent-block Parcel }
                        \RIDLCONT
                        \RuleVar{ Lot} \\
                      & \RuleVar{elementary-surveyed-unit-of Parcel} \\
                      & \RuleVar{has-road-area-related-to} \mbox{ Current-App} \\
                      & \RIDLAND \\
                      & \RuleVar{received-on Date} < 
                        \RuleVar{Date marks-receipt-of} \mbox{ Current-App} \\
                      & \RIDLAND \\
                      & \RuleVar{ received-on Date} \geq \RuleVar{Date marks-receipt-of}
                             \mbox{ Current-App} - \RuleVar{2 years)}
   \end{array} \]

\begin{figure}[htb]
\centering
\epsfig{figure=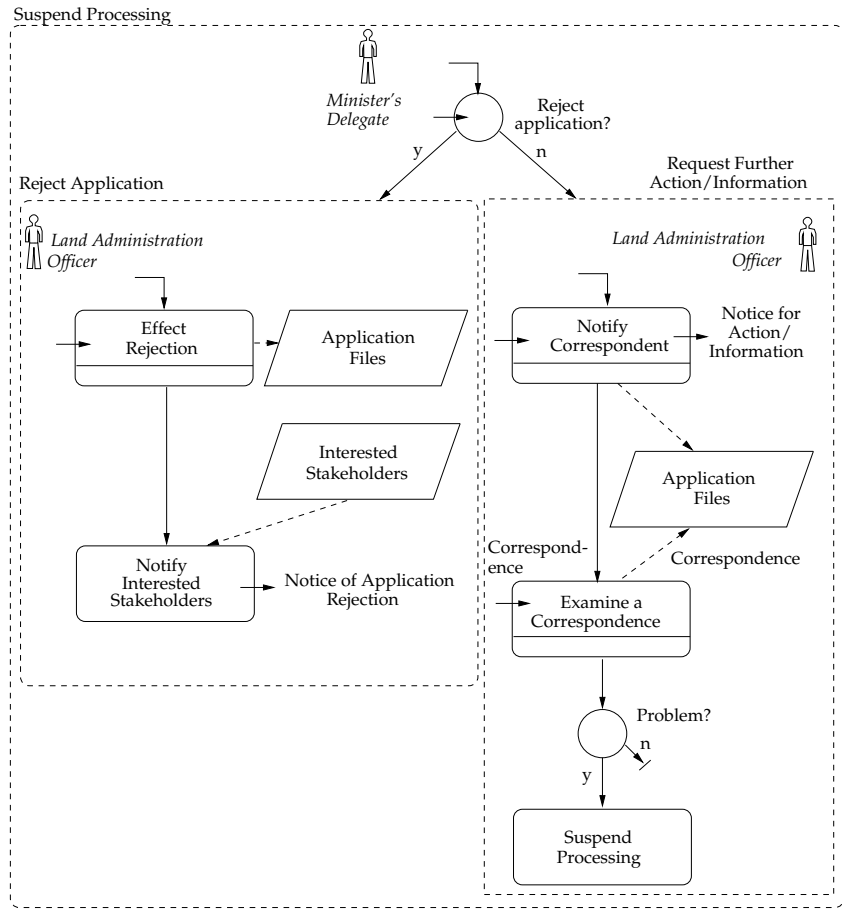,width=0.7\linewidth}
\caption{An example of recursive decomposition}
\label{suspend.fig}
\end{figure}

Like a simple decision, the result of a complex decision is either 
a positive or negative outcome. In this example, a negative outcome 
results in the 
execution of \textsf{Suspend Processing} which is depicted in 
Figure~\ref{suspend.fig}. In brief,
it either results in a rejection of the \SF{Application} or a request for
further information, both of which result in the appropriate outgoing
notifications sent to the \SF{Interested Stakeholders}. Rejecting an 
\textsf{Application} not only means
updating the state of object, but ultimately terminating the workflow
associated with the \SF{Application} instance. In section
section~\ref{errorhandlingSec}, the subject of our treatement of the
\EM{\ErrorHandling\ Principle}, 
we discuss how this type of operational error and other types, are handled.
An interesting feature of this example is the use of \EM{recursive}
decomposition, i.e.\ \SF{Suspend Processing} invokes itself.
Unlike a pure structured decomposition as adopted in
structured process modeling techniques (e.g.\ Data Flow Diagrams), the
control-flow nature of this technique permits this.

\begin{figure}[htb]
\centering
\epsfig{figure=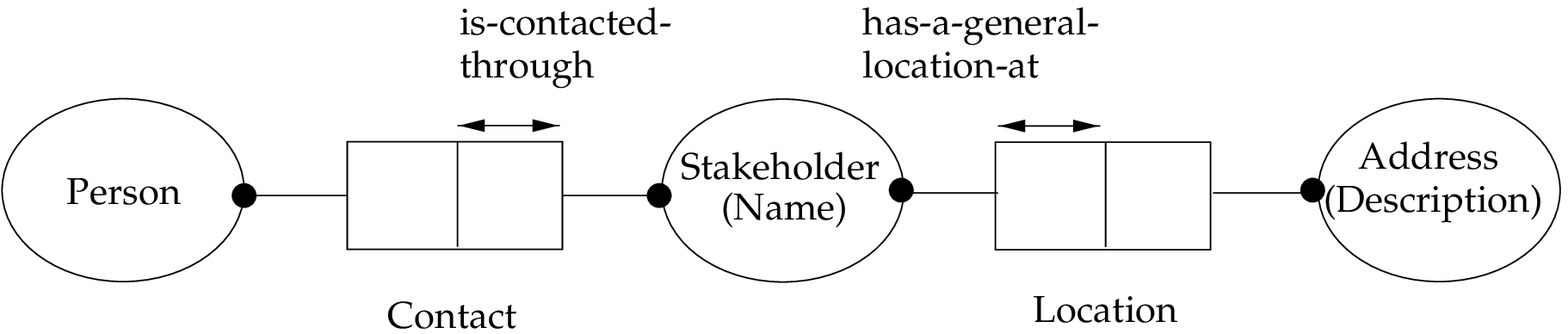,width=0.7\linewidth}
\caption{Schema associated with the \textsf{Candidate Stakeholders} message}
\label{can.stakeholder.fig}
\end{figure}

Returning to \textsf{Initial review passed?} in Figure~\ref{app.invest.fig},
its positive outcome results in the \textsf{Preparation} for a detailed 
investigation of the \textsf{Application}. This involves publication of the road
closure intention in the Government Gazette; done through the \SF{Gazettal service} 
(external to the department). The department seeks the views of stakeholders 
if it is required to do so. This is done through \SF{Seek Views}, depicted in 
Figure~\ref{seek.fig}.
First the \SF{Candidate Stakeholders} need to be determined. These are
obtained through \SF{Parcel Info} (an external service which
accesses a cadastral database identifying the surrounding parcels, 
utilities etc.). Then the contents of the message needs to be inserted 
into the \SF{Candidate Stakeholders} object store. 
As an alternative to performing this through a HCI, as would ordinarily be
the case, it is possible to automate the message transfer. This requires
that the target object store schema and the source message schema both be
compatible. The message's schema is illustrated in
Figure~\ref{can.stakeholder.fig}.

Using the fact type denotations, \SF{Contact} and 
\SF{Location}, both of which also apply for the \SF{Candidate Stakeholders}
object store, the following LISA-D statement illustrates the automatic
message transfer:

\[ \arraycolsep = 0pt
   \begin{array}{ll}
   \RIDLADD \RuleVar{ Contact } & \RIDLIN \RuleVar{ Candidate Stakeholders~}
            \RIDLTO \RuleVar { Contact} \\
   \RIDLADD \RuleVar{ Location } & \RIDLIN \RuleVar{  Candidate Stakeholders~}
                               \RIDLTO \RuleVar { Location}
   \end{array} \]

A \textsf{Notice of Road Closure} is then sent to each \textsf{Stakeholder},
illustrating how \EM{bulk} messaging (the analogue of bulk database updates)
can be incorporated in LISA-D:

\[ \arraycolsep = 0pt
   \begin{array}{ll}
   \RIDLSEND & \RuleVar{ Notice of the Road Closure } \RIDLTO 
   \RIDLEACH \RuleVar{ Person is-contact-for Stakeholder} \mbox { s} \\
   & \RIDLAT \RuleVar{ Address is-contact-for Stakeholder} \mbox{ s} \\
   \end{array} \]

Now the sending of the messages is required to occur no later than one 
day after the date of gazettal. This illustrates the need for a 
\EM{temporal constraint} in the postcondition in \textsf{Seek Views}:

\[ \arraycolsep = 0pt
   \begin{array}{ll}
   \ENDDATE \BF{(} \RuleVar{Seek Views} \BF{)} \leq 
\RuleVar{Date is-gazetted-date-of Application} 
\mbox{ Curr-Application} + 1
   \end{array} \]

As an example of a temporal constraint on preconditions, a \SF{Site Inspection}
is not allowed to occur more than prior to two months before the intention
for road closure has been ``gazetted'':
\[ \arraycolsep = 0pt
   \begin{array}{ll}
   \STARTDATE \BF{(} \RuleVar{Road Inspection} \BF{)} \geq 
\RuleVar{Date is-gazetted-date-of Application} \mbox{ Current-App} + \mbox{2 months}
   \end{array} \]

\STARTDATE\ and \ENDDATE\ indicate the need for temporal functions which 
provide the start and end dates (times) of process object execution. This 
implies 
that certain execution
statistics about process objects should be maintained. This
allows time durations to also be used within constraints, for example, for
``timeouts''. Also process execution dependency can further be qualified
through temporal constraints. For example: run a number of processes at 
some time, simultaneously (parallelism); or within a time duration of each other 
(sequence); run a process repeatedly within a certain time period
or cyclically at time points (repetition). 
Such constraints can apply to messaging as well, e.g.\ contingent service 
access for process objects if
messages have not returned within certain times.

\subsection{Service Model}
\label{service}

So far features of \KerName's object and process modelling techniques
have been presented. In this section, the last and most pivotal of the models, 
the service model is described. The service modelling technique of \KerName\ 
is entirely a new proposal. It allows business services to be explicitly 
modelled such that, in accordance with the \EM{\InfoHiding\ Principle}, service 
requests are insulated from the resultant business processing. Recall (from 
section~\ref{infohiding}) that the key determinant of what business processing 
is required for a given service request, is the state of the service. In turn, 
a particular service state is dependent on the success of the elapsed business 
processing. Of course, the states of a service should describe the \EM{lifecycle} 
of a service in a way which is meaningful to its stakeholders, and should not
be used as a business processing ``log''.  

It can be seen that a convenient way to model a service is an object. Objects,
afterall, encapsulate processes, and their behaviour is described through a 
lifecycle of states and state transitions (see e.g.\
\cite{Book:91:Rumbaugh:OO, Book:88:Shlaer:OO}).
That is, for a given state, an object reacts to a set of events through the 
activation of actions, possibly if certain conditions are satisfied. 
Applied to business services, service requests are specified as events 
while the actions represent business processes. In general, events represent 
occurrences which signify changes of state in the business scope. 
Examples of events include the receipt of a message, changes in time and the
termination of process/decision and the occurrence of an abort.
A business service's state-dependent reaction to an event includes the
triggering of a process model object. Process model objects, of course,
trigger other such objects, produce messages or effect updates and 
retrievals from object stores. As the workflow progresses, different events 
are raised, and again, the service object may react to these, further 
propagating processing. 

It can be seen that in \KerName, the \EM{state-centric} modelling
cognition of the service model complements the \EM{process-centric}
modelling cognition of the process-model. This, in effect, represents a
declaritive versus imperative appropriation in specifications to effectively
capture business processing semantics. That is to say, the higher
levels of business processing are described declaritively through 
service models wheras at lower levels are decribed (more) imperatively 
through process models. The exclusive adoption of one dynamic modelling
paradigm is considered unsuitable. A complete state-centricity
leads to complete object-orientation which has the disadvantage of turning 
each process model object into  an ``island'' specification. On the other hand, 
a complete process-centricity would result in complex business service 
specifications, given the permutations of exceptions of process invocations 
for the different events.

\begin{figure}[htb]
\centering
\epsfig{figure=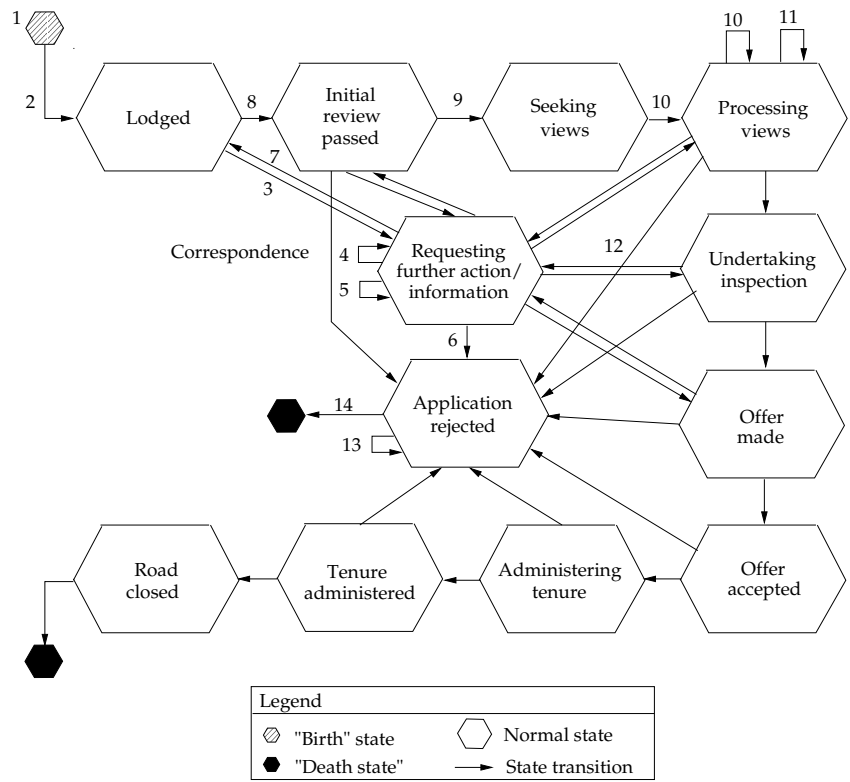,width=0.7\linewidth}
\caption{Service Model for Road Closures}
\label{roadcl.service.fig}
\end{figure}

The service model for the road closures business transaction is depicted in
Figure~\ref{roadcl.service.fig}. The service model consists of: normal states 
(large polygons), e.g.\ \textsf{Lodged}, \textsf{Initial review passed} and 
\textsf{Application rejected}; special states indicating the ``birth'' of a 
service (small unshaded polygon) and the ``death'' of a service (small 
black-shaded polygon); and state transitions (arcs). The special states allow 
transitions to the first possible state(s) (\textsf{Lodged}) and last possible 
state(s) (\textsf{Application rejected} and \textsf{Title issued}) to be 
specified within the same service context (otherwise some global service
would be required to undertake these). Of course, this issue relates to
the suitability and comprehensibility rather than the expressive power of 
the technique.

The event-condition-action (ECA) paradigm which has been adopted for 
active rule specification in database systems, e.g.\
\cite{Article:90:Chakravarthy:ActDB},
and conceptual specification languages, e.g.\ 
\cite{Article:91:Loucopoulos:Tempora}, is adapted for event specifications. 
These, of course, are attached to the service object state transitions. 

The first event is the arrival of the message \textsf{Application Documents}.
This is an example of a \EM{messaging event}. 
It is distinguished from an actual external event e.g.\ the signing of a 
contract for an estate development. In general, the inclusion of such
external events do not seem necessary for workflow specifications, although
they could be captured through an event dependency formalism separate to the
workflow specification. This first
event leads to the service object (instance) creation in the 
\textsf{``birth state''}. Upon this creation, it sends the message to 
\textsf{Application Entry} thereby triggering the workflow described in 
Figure~\ref{app.lodge.fig}:

\[ \arraycolsep = 0pt
   \begin{array}{ll}
     & \WHEN \MSGFROM{Application Documents} \\
     & \THEN \MSGTOPROCESS{Application Documents}{Application Lodgement}
   \end{array} \]

The transition to the \textsf{Lodged} state occurs when the \textsf{Application}
object is first entered into the \textsf{Application Database} (2) - an example 
of a \EM{database state event}. The expression is formulated using LISA-D
demonstrating how conceptual data specification
languages can be used by ECA languages. The predicate, in
this case, is an arbitrary fact type with a mandatory role since this will
evaluate to ``true'' after the \textsf{Application} is stored in the database. 
No \THEN part follows since the workflow execution still continues (without
the need for invoking further processing). This further illustrates the 
importance of service states capturing the required perceptions of service 
stakeholders, independent of the underlying workflow execution:
 
\[ \arraycolsep = 0pt
   \begin{array}{ll}
 & \WHEN \RuleVar{Application received-on Date} \\
   \end{array} \]

A problem may be found in the \textsf{Application} (this relates to the
internal checks done as part of the decision \textsf{Initial review
passed?} described in Figure~\ref{review.fig}). In this case, a 
\textsf{Request Further Action/ Information} may executed. Its issue of
a \textsf{Notice of Further Action/Information} message - a messaging
event but this time outgoing - is detected by the service object for the 
next state transition (3):

\[ \arraycolsep = 0pt
   \begin{array}{ll}
     & \WHEN \MSGTO{Notice for Further Action/Information} \\
   \end{array} \]

The subsequent \textsf{Correspondence}, like all incoming messages 
from the environment, is sent to the service object, and so the service 
object further activates processing. The \textsf{Correspondence} should
be examined, and so no state change results (4):

\[ \arraycolsep = 0pt
   \begin{array}{ll}
     & \WHEN \MSGFROM{Correspondence} \\ 
     & \THEN \MSGTOPROCESS{Correspondence}{Examine a Correspondence} \\
   \end{array} \]

Although not included in the process model description, 
\textsf{Correspondence} may not be satisfactory in which
case the \textsf{Suspend Processing}
may be reinvoked with the \textsf{Minister's Delegate's} decision to
\textsf{Reject Application?} - a \EM{processing event} involving negative
decision termination:

\[ \arraycolsep = 0pt
   \begin{array}{ll}
     & \WHEN \DECISIONEVENT{Reject Application?}{REJECTED} \\ 
   \end{array} \]

or a positive decision termination resulting in the transition to
the \textsf{Application rejected} state (6):

\[ \arraycolsep = 0pt
   \begin{array}{ll}
     & \WHEN \DECISIONEVENT{Reject Application?}{ACCEPTED} \\ 
   \end{array} \]

The above ECA rules provide some indication of the types of
events which a service can react to. In the full case study, the need for
the following further  events types was identified, e.g. when: 
process objects 
commence execution; process objects fail to commence execution over 
a range of times; terminating aborts occur.
Also for temporal constraints, we have identified the need for
the \STARTDATE\ and \STARTTIME\ of service states, the \SENDDATE, 
\SENDTIME, \RECDATE\ and \RECTIME\ of messages.
(Recall the \STARTDATE, \STARTTIME, \ENDDATE\ and 
\ENDTIME\ of processes are also required to be known).
Nevertheless, we recognise that the expressive completeness
and suitability of service ECA language are an open issue.

\subsection{Exception modelling}
\label{errorhandlingSec}

\begin{figure}[htb]
\centering
\epsfig{figure=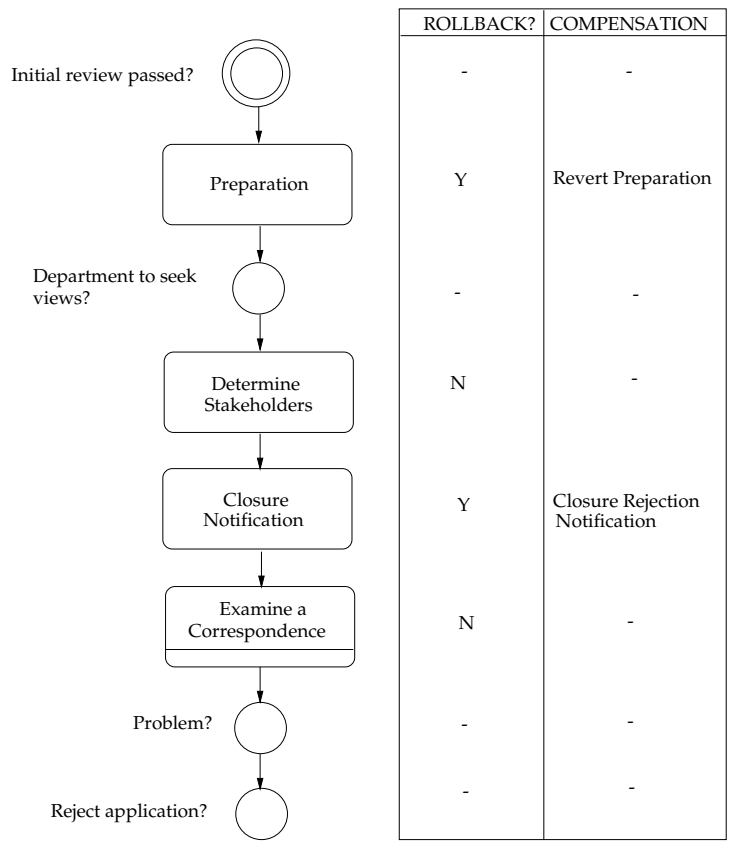,width=0.6\linewidth}
\caption{Rollback recovery example for road closures}
\label{roadcl.rollback.fig}
\end{figure}

In Figure~\ref{roadcl.rollback.fig}, an example of a rollback recovery
is depicted for road closures. In it, the workflow has progressed
to the point where an abort has been raised to reject the
\SF{Application} during \EM{Process Views}. The elapsed 
workflow includes the initial investigation and
the preparation, the seeking of views and the processing of one incoming
and problematic view (e.g.\ a stakeholder's rejection of the application is
not reconcilable). A rollback therefore has to occur. The rollback 
specification of each process object is presented in the table. It
indicates whether a rollback is required, and if so whether a particular 
compensation applies. 

From the table, it appears that decisions do not require rollbacks. This
seems intuitively acceptable since decisions are simple processes. In a
complex decision having a terminating abort, other decisions are merely
terminated. Also, of course, when no rollback is to be performed, no
compensation applies. However, when a rollback is required, a compensation
may apply (when the process is not in the current commit
grain and therefore when an undo cannot be performed). In this example, the
rollback strategy is running \SF{Closure Rejection Notification} involving 
messaging containing corrective information and \SF{Revert Preparation} 
involving an 
idempotent database update. This illustrates how error handling 
can be localised into process objects as opposed to the traditional approach 
which centralises error handling into a single routine; clearly a gain in
comprehensibility and suitability.

The modelling of rollforward recovery is not discussed.

\section{An integrated conceptual modelling kernel}
\label{formal}

Following the motivation and illustration of \KerName's modelling 
concepts and techniques, its definition including a formal syntax is now 
provided. In Section~\ref{formal-scope}, the main concepts are described and 
placed into the context of the business scope. The process modelling 
technique is described in section~\ref{pm-tech} while the service 
modelling technique is described in section~\ref{sm-tech}. The object 
modelling technique has not been changed from that described in
\cite{Report:95:Proper:CDMKernel},
and so no definition is provided here.
Axioms for inter-model consistency are not provided, nor is a definition of
the formal semantics. Both are the subject of future research.

\subsection{Business Scope}
\label{formal-scope}

In its broadest sense, \KerName\ provides a set of kernel techniques which
allow the domains of business transaction workflows to be modelled in an
integrated fashion. The techniques involve object, process and service
modelling. The notion of \EM{kernel} means that the techniques are not 
necessarily concrete techniques, although they may be applied as such. 
Rather they provide a set of \EM{abstract} concepts and features from 
which concrete techniques may be specialised and further developed.

\begin{figure}[htb]
\centering
\epsfig{figure=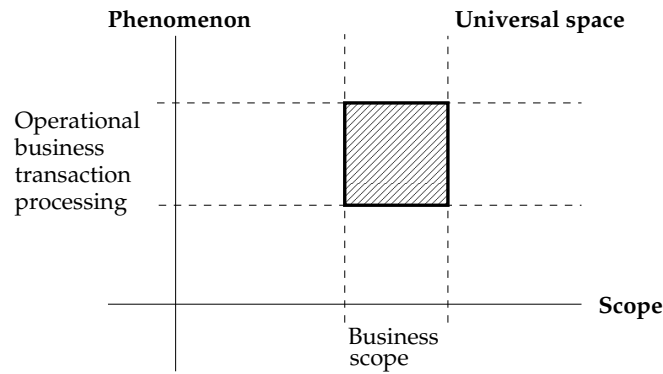,width=0.55\linewidth}
\caption{Framing \KerName's concepts within universal space}
\label{universe.fig}
\end{figure}

At the outset, we consider it important to establish the various 
\EM{scopes} from which the concepts belong. This serves to discern
a technique's highest level of perception of a domain; an important 
factor during the initial phases of analysis. In other words, the basis 
for \EM{organisational embedding} should be clear.
For \KerName, the highest possible scope is a business scope.
It is partitioned into a business domain, which is the focus of 
the modelling, and the business environment, which is required 
so that the business domain's external interaction may be modelled. 
A business scope is, in itself, a concept, aggregrating
all other concepts. It exists within an absolute scope - 
a universal space say - which may contain other types of scopes.
The set of concepts which are relevant to
\KerName, are not only bounded ``vertically'' by a business scope
and all the possible concepts relevant to a business scope, but
also ``horizontally'' to type of phenonemon of relvance. This, of
course, is that type processing which we have characterised as 
operational business transaction workflows.
As depicted in Figure~\ref{universe.fig}, the intersection of this 
``vertical'' and ``horizontal'' universal space represents 
the set $\Concepts$ of concepts relevant to \KerName.

The concepts that we begin with are those which assist in
\KerName's organisational embedding. In other words, these
concepts are enterprise (or business) concepts which may be
refined into further concepts within the specific technqiues.
As an example, the enterprise concept process is refined into
process, decision and synchroniser in the processing modelling
technique.
At the same time, it should be understood that there is no
fundamental difference in the concepts used for the enterprise 
and detailed IS modelling, as in, for example, 
\cite{PhdThesis:93:Ramackers:BehaviourMod}. Rather we follow
an observation made by the FRISCO Group
\cite{Report:96:Falkenberg:ConcFramework},
that ISs are organisational (sub)systems, and so the conceptual 
basis for IS modelling is also relevant for organisational 
(sub)systems; and an IS implementation platform, 
workflows reduce the traditional gaps of their respective 
cognitions.

A convenient mechanism for determining the set of concepts 
which belong in a business scope, a business domain and a business 
environment, is the set of organisational units
$\OrgUnits \subseteq \Concepts$.
This is because organisational units are the means by which processing
and storage entities are dispersed within an organisation.
This is defined through the relation 
$\Structure \subseteq \OrgUnits \times \Concepts$. 
An organisational hierarchy is structured as a rooted directed acyclic 
graph of organisational units where the ``top'' of each graph is an 
organisation. The relation $\SubOf \subseteq \OrgUnits \times \OrgUnits$ 
allows parts of a number of organisations to be pertinent for a workflow, 
and so represents a ``forest''. As scopes over the ``forest'', a business 
scope $\BusScope$, a business domain $\BusDomain$ and a business 
environment $\BusEnv$ are all subsets of business concepts. This is 
depicted in Figure~\ref{bus.scope.fig}.


The degree of workflow modelling in a business domain is whole while
that in the business environment is only partial. 
This is alluded to by the \EM{\InfoHiding\ Principle} which aims to 
encapsulate workflow specifications where possible.
Of course, some concepts may be used in both the business domain and the
business environment.
For this, a concept qualified as \EM{internal} means that it
exists in the business domain while that qualified as \EM{external}
means that it exists in the business environment.

The remaining types of concepts are:
\begin{enumerate}

\item A set $\Actors$ of actors. An actor is instrumental in 
undertaking business transactions. Actors, as such, do not undertake the 
processing, but (as described below) do so in roles. 
Only internal actors are of relevance{\footnote{This is philosophical rather
than fundamental. For example, this view would be not be acceptable 
for techniques adopting actor-centric paradigms}.}.

\item A set $\Roles$ of roles. A role is a particular functional charter 
required for one or more business transactions. Actors are assigned to roles. 
This is given through the relation $\Assign \subseteq \Actors \times 
\Roles)$. As with actors, the only roles of relevance are 
the internal ones. 

\item A set $\Processes$ of processes. Processes are \EM{prescriptive} units of 
functionality which allow business transaction processing, among others, 
to be described. Typically, processes involve human(manual) actions, 
computerisable actions such as reading, writing and updating data and
undertaking dialogues whether human or computerised.
As directly apparent from the \EM{\InfoHiding\ Principle}, the only processes
of relevance are the internal ones.

Actor roles are required for to undertake processes. This is described by the 
relation $\Undertake \subseteq \Roles \times \Processes)$. 
Moreover, actors should only be allocated to roles required by 
processes which are in organisational units that the actors are assigned 
to. This is defined using the following axiom, noting that $\circ$
represents a relation composition:

\[\Assign\circ\Undertake\circ\Structure \subseteq \Structure \]

\item A set $\Services$ of services. Services are \EM{descriptive} units 
of functionality which allow business transaction processing, among others,
to be described. In fact, a service represents a mechanism by which 
business transactions are accessed, without, 
as described by the \EM{\InfoHiding\ Principle}, knowledge of how the 
business transactions are undertaken. It follows that both internal
and external services are of relevance.

\item A set $\ObjTypes$ of object types. Object types are either 
informational or material. An informational object, e.g.\SF{Application} 
is abstract having static (but not behavioural) properties. A material 
object, e.g.\ \SF{Application Files}, is tangible having no further 
properties of interest. The only object types of relevance are the internal 
ones.

Although we have not provided
the definition of \KerName's object modelling technique, it goes without
saying that an object belongs to a schema definition(s). This is given 
by the function $\Schema\colon \ObjTypes \Func \Powerset^+(\SchNames)$,
where $\SchNames$ is the set of schema names.

\item A set $\ObjStores$ of object stores. Object stores provide a 
persistant, structured storage for objects. Of course, they may store 
material object types, e.g. a car pool, or informational object types, 
e.g. a paper general ledger or a database, but not both. 
Some object stores are distributed e.g. distributed databases, and so
may reside in a number of fragments. This is given 
by the partial function $\Fragment\colon \ObjStores \PartFunc 
\FragmentName$ where $\FragmentName$ is the name of the fragment.

The only object stores of relevance are the internal ones.
Note, the fragments of a distributed database which lie outside 
the boundary of an organisation, are considered to still be part of 
a business domain. This is because a distributed database is a single 
logical entity. This is not true of a federation of databases accessed 
by a given process. Any database in the set may be autonomous and may 
reside outside the business domain. Access to any such external database 
should occur through services. 

\item A set $\Messages$ of message types. Messages are used to transfer 
data during processing. They may contain structured data (objects) or 
unstructured data. Both internal and external messages of relevance; 
the relevance of external messages follows that of external services.

\item A set $\MesBuffers$ of message buffers. Message buffers provide a
transient, unstructured storage of messages, e.g.\ an in-tray, an
electronic mail box or a set of ``pigeon holes''. Message buffers
may be allocated to particular message types, given by the relation
$\MesAlloc \subseteq \MesBuffers \times \Messages$.

The storage (retrieval) of messages to (from) a message buffer follows
a message protocol. This is given by the function 
$\MProtocol\colon \MesBuffers \Func \MPNames$, where $\MPNames$ is
the set of message protocol names. Instances
of it include FIFO and FILO queues, a random order or an order
specified by a predicate.
\item A set $\Events$ of events. An event is a discrete and instantaneous 
occurrence representing a change of state within the business scope. In its
broadest sense, events result from: interactions between
processeses producing, or produced during, business transaction 
execution; changes in time and changes in object states (as a result of database 
updates). Interactions are discussed in more detail in 
section~\ref{pm-tech}. 

It follows that the events which a domain can perceive are internal to the 
domain, since these result from interactions on processes (recall,
only internal). 
So, for example, the event of an arrival of an \textsf{Application} and not
its caused event of the establishing of a contract for the
development of an estate, is perceived by the Road Closures domain. 
\end{enumerate}

In general, the name of a concept is determined through the function
$\Name\colon \Concepts \Func \ConcNames$, where $\ConcNames$ is the set
of names.

\subsection{Process modelling technique}
\label{pm-tech}

The process modelling technique, allows the prescriptive aspects 
of business transaction processing to be modelled. This means that 
processing which results directly or indirectly from service 
requests. In its broadest sense, a process model consists of a 
set $\PEntities$ of processing entities, a set $\SEntities$
of storage entities and their interactions between these which
characterise the modelling of the processing.

\subsubsection*{Processing entities and interactions}

\begin{figure}[htb]
\centering
\epsfig{figure=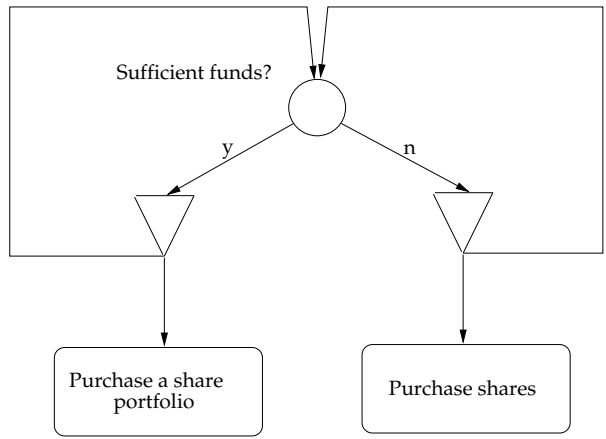,width=0.5\linewidth}
\caption{Implementing a ``fork'' using a synchroniser}
\label{fork.fig}
\end{figure}

$\PEntities$ generalises not only processes (which are visible at
the business level) but also concepts which refine processes,
namely decisions and synchronisers. In fact, $\Processes$,
$\Decisions$ and $\Syncs$ partition $\PEntities$.  
Decisions allow ``moments'' of processing uncertainty where a 
(possibly non-deterministic) choice of execution paths may be 
followed, depending on the outcome of the decision.  
Synchronisers allow ``moments'' of synchronisation. A common
example of synchronisation, like that depicted in
Figure~\ref{roadcl.process.fig}, requires a number of execution
paths to be reached, prior to further processing being executed.
A further application of a synchroniser is depicted by the example
of a share purchasing process of Figure~\ref{fork.fig}. 
Here, a decision is made to determine
whether \SF{Sufficient funds?} (from a ``dynamically'' 
updated account) are available to
\SF{Purchase a share portfolio} (shares from a number of
types). If not, basic shares are purchased, i.e.\ \SF{Purchase
shares}. In either case, when the processes are ``forked'' off for
the purchase, control is returned to the decision to ensure
that the purchasing process is continuous. 

Interactions between processing entities are specialised as 
\EM{triggering} and \EM{messaging}.
Triggering serves to activate execution, i.e.\ control flow.
This is defined through the relation $\Trig \subseteq 
\PEntities \times \PEntities$, where 
$x \Trig y$ means that $x$ is triggered by $y$. 

\begin{figure}[htb]
\centering
\epsfig{figure=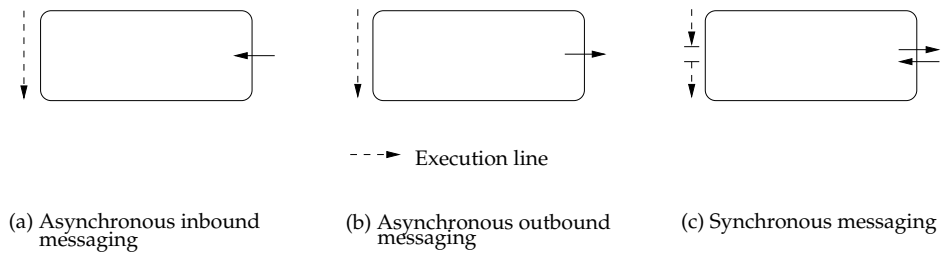,width=0.8\linewidth}
\caption{Messaging modes}
\label{messaging.fig}
\end{figure}

Messaging, as (just) described above, serves as a 
communication mechanism, i.e.\ data flow. 
In particular, it involves the sending or receiving of messages, 
which while having some effect on execution control, is an issue
orthogonal to it.
Only processes and decisions may be involved in messaging, hence 
the definition of the set of non-synchronisers 
$\NonSyncs = \PEntities - \Syncs$. 
Figure~\ref{messaging.fig} illustrates the different messaging
modes. 

Asynchronous messaging involves the sending \SF{(a)} or
or receiving \SF{(b)} of a message at some stage during the execution
of a processing entity afterwhich execution continues. 
Messaging may also be synchronous in which case a message is sent and 
execution is suspended until a message is received \SF{(c)}. Note,
any messaging involving a message receipt first, is automatically
asynchronous since no suspension of execution relates to it. In
other words, synchronous messaging involves message sending first
only.

\begin{figure}[htb]
\centering
\epsfig{figure=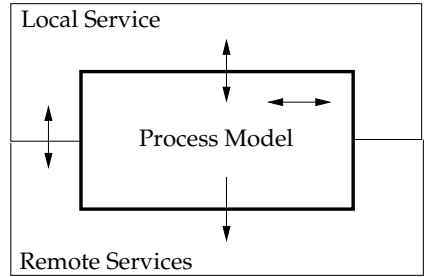,width=0.35\linewidth}
\caption{Scope of messaging interaction}
\label{messaging.scope.fig}
\end{figure}

Three disjoint subsets of non-synchronisers involved in messaging are 
defined:     
the set $\AINonSyncs$ involving asynchronous inbound messaging, the set
$\AONonSyncs$ involving asynchronous outbound messaging and the set
$\SNonSyncs$ involving synchronous messaging.
The function
$\Message\colon $\AINonSyncs \union $\AONonSyncs \union \SNonSyncs
\Func \Powerset^+(\Messages)$ yeilds the messages involved; the power
set signifies the fact the multiple messages may be
involved. 
Moreover, messages are received either from the service local to
the non-synchroniser or from another non-synchroniser in the
same process model. In other words, messages received from
the environment are always received via the local service.
Messages may be sent to other non-synchronisers in the process
model, the local service or remote services. 
Figure~\ref{messaging.scope.fig} depicts this
scope of messaging interaction. The function $\Mesg\colon
\PEntities \union \Services \Func\PEntities \union \Services$
describes the messaging interaction.

(??? Interaction points which have dialogue specifications and
protocol - and types HH, CC and HC ??? Should structural aspect of
dialogue specifications have a schema defintion much like
``view'' objects).

\subsubsection*{Decomposition}

A feature of process models is decomposition where processes and
decisions may be refined in detail. Hydra allowed processes to be
decomposed, however the decomposition of a decision, 
i.e.\ a \EM{complex} decision, is a new proposal. We illustrated
complex decisions in Figure~\ref{review.fig} and 
Figure~\ref{compdecis.fig}.  

In general, decompositions allow parts of a process 
model to be modularised and reused. For this the partial function
$\Sup\colon \PEntities \PartFunc \ConcNames$ is defined 
to provide the name of the decomposition that the processing
entity belongs to. If $\Sup(x) = v$, this means that processing 
entity $x$ is part of the decomposition of $v$. Of course, the
names of decisions and processes at the same level of decomposition
should not be the same:
\[ x \in \Processes \land y \in \Decisions \implies 
\Name(x) \neq \Name(y) \]
Also, unlike a process decomposition, a decision decomposition 
is only permitted to have decisions:
\[ \Sup(x) = v \land \Ex{d \in \ConcNames}{\Name(d) = v} 
\implies x \in \Decisions \]

So that processing entities are ``rooted'', a process model is required 
to have a unique process at the top of the decomposition heirarchy:
\[ \Eu{p \in \Processes}{\isundefined{\Sup(t)}} \]
Within each decomposition, a process model is required to have a
set of initially executed entities (if more than one,
parallel execution follows). For this, the 
partial function $\Init \subseteq \Sup \intersect \PEntities$ 
is defined.

Decomposition hierarchies do not have a cyclic
constraint (e.g.\ the ``downwards'' only decomposition of data
flow diagrams), but may include recursive decomposition. 
Figure~\ref{suspend.fig} contained an example of this.
Two contraints apply to interactions across decompositions. 
Firstly, triggers should not cross decomposition boundaries:
\[ x_1 \Trig x_2 \implies \Sup(x_1) = \Sup(x_2) \]
Secondly and similarly, intra-service messaging should not 
cross decomposition boundaries:
\[ x_1 \Mesg x_2 \implies \Sup(x_1) = \Sup(x_2) \]
where $x_1,x_2 \in \NonSyncs$.

\subsubsection*{Storage entities}

Storage entities $\SEntities$ are used to generalise messages,
message stores and object stores. $\Messages$, $\MesBuffers$ and
$\ObjStores$ partition $\SEntities$. A schema provides a definition, 
pertinentto the storage entities. For messages and object stores, it
represents the information grammar. 
For message buffers, it incorporates characteristics about the message
stores, e.g.\ types of messages permissible, message quantities.
A partial function, $\Schema\colon \SEntities
\PartFunc \SchNames$, provides the schema definition. 
   
Of course, processes and decisions access data through storage 
entities. Following from the above discussion,
\EM{direct} storage entity access implies storage entities within 
the same level of decomposition. For convenience, the \EM{decomposition
names} are distinguished from \ConcNames\ as those having at least one
processing entity in them:

\[ \ConcNames_d =
   \Set{v \in \ConcNames}
       {\Ex{x \in \PEntities}
           {\Sup(x) = v}
       } \]

Storage entities can then be assigned to decompositions having at 
least one processing entity through the function $\Locse\colon 
\ConcNames_d \Func \Powerset(Var)$.
To assist in the expressive power of the processing, local variables 
may also be defined within a decomposition, hence the function 
$\Locvar\colon\ConcNames_d \Func \Powerset(\Varclass)$.
Processes and decisions may access those storage entities and variables
within their level of decomposition and lower. They cannot access those 
in higher levels of decomposition. Hence, storage entities and variables
have a scope in the same sense as scopes of variables in programming 
languages such as ALGOL
(see e.g.~\cite{Book:76:Wijngaarden:algol68}).
To avoid naming conflicts, the names
of local variables and storage entities in that decomposition should
differ, i.e.\ $\Locvar(v) \intersect \Locse(v) = \emptyset$.
Also, processes may reference and change data but decisions are only
permitted reference (this is formalised in the respective sections below).

In addition to variables, data may be passed between processes through 
message buffers. 
Messages may be consumed from, or produced into, a message store. This 
is defined through the 
functions $\Consume\colon \Processes \Func \Powerset(\Varclass)$ 
and $\Produce\colon \Processes \Func \Powerset(\Varclass)$ respectively.
If $v \in \Consume(p)$, then process $p$
consumes from message store $v$.
Note that if a process $p$ does not consume
from any message store, then $\Consume(p) = \emptyset$.
Processes may only consume from, and produce for, message
stores which are part of the same decomposition:
\[ \Consume(p) \union \Produce(p) \subseteq \Locse(\Sup(p)) \]

\subsubsection*{Processes}

The assignments, reads, writes and updates
which typify process specifications are expressed in LISA-D. In Hydra, a 
sequence of these may be assigned into a LISA-D transaction classified 
as the LISA-D syntactic category $\Trans$. This should be extended to include
\KerName's message sends oustide the decomposition scope (as distinct from 
producing messages in message stores within the decomposition).
Since processes may be composed, only atomic processes are assigned a
LISA-D transaction, i.e.\ 
$\TransAssign\colon \ConcNames\backslash\ConcNames_d \Func \Trans$.

Sometimes, processes should not be able to start or terminate when certain 
conditions are not fulfilled. These conditions can be expressed by means
of pre-condition and postcondition. Pre- and postconditions are expressed 
by means of LISA-D predicates which in Hydra are classified into the
LISA-D $\Predicate$ category. Hence the functions
$\Pre\colon \Func \Predicate$ and $\Post\colon \Func \Predicate$
respectively.
The predicates may reference the names of storage entities and local 
variables. Also the special LISA-D predicate \textsf{true} can be
used to indicate that there is no pre- and post-condition.

\EM{Exclusive processes} (called transaction tasks in Hydra) are processes 
which run isolation, i.e. no other process can run when an exclusive
process is running. Hence the set $\ExcProcesses \subseteq \Processes$. 
The notion of exclusive processing should not be confused with 
exclusive locking which does allow a concurrent
execution of procceses albeit that an exclusive lock on an object 
\EM{blocks} any other lock until the exclusive lock is released.
Typical examples of exclusive processes are systems maintenance jobs
involving database checkpoints and systems backups, or systems
configuration involving installations and upgrades. Prior to starting an 
exclusive process, all active processing entities should be \EM{quiesced}. 
This, of course, is an issue of formal semantics.

\subsubsection*{Decisions}

To allow ``moments'' of uncertainty to be specified, decisions
have output triggers - one for a postive outcome and one for a 
negative outcome - which are are assigned LISA-D predicates. These
decision rules can be compared to \EM{guarded commands}
as introduced in~\cite{Article:75:Dijkstra:GuardedCommands}.
As with guarded commands the decision rules of a decision are not 
necessarily disjoint. Therefore, nondeterministic choices
can be modelled.

The predicates may refer to values of variables and object instances of
object stores and messages. Synchronous and asynchronous messaging is 
allowed by decisions so that object instances of messages may be 
referenced.
The output triggers of a decision may terminate, hence the 
subset $\Decisions_t \subseteq \Decisions$. Some decisions signal an
abort message which signify the intention to
terminate a complex decision, hence the subset $\Decisions_t{_a}
\subseteq \Decisions_t$. For those decisions which do not terminate,
a processing entity can only be triggered by a decision if
a trigger exists from that decision to that processing entity while the 
associated decision rule evaluates to true.

Decision rules are recorded by the function:

\[ \Drule\colon
       ((\Decisions \times \PEntities)
        \intersect \Trig) \union
        (\Decisions_t \times \set{\multimap}) \Func
       \Predicate
\]
$\Drule(k,\multimap) = d$
means that $k$ is a terminating decision that may
lead to termination if $d$ is fulfilled.

\subsubsection*{Recovery}

To fulfill the \EM{\ErrorHandling\ Principle}, the detection of
operational errors are signalled by two forms of abort messages.
The first is a failure abort, which as the name suggests,
results from failures such as system crashes. The second is non-failure
aborts which are (deliberately) produced within a
business transaction e.g.\ by a process, decision or service. Note,
aborts used to terminate decisions are not a type of non-failure
aborts. Their function is more restrictive.
The denotation of these aborts is described in $\set{\mbox{`\SF{F-Abort}'},
\mbox{`\SF{NF-Abort}'}} \subseteq \Name(\Messages)$.

As described in section~\ref{errorhandling}, the recovery strategy for the 
errors involving failure aborts is a \EM{rollforward
recovery} whereby a \EM{redo} operation is applied
to a ``crashed'' processing entity. If the processing entity does not
start after numbers of restart attempts, after time periods or
after combinations thereof, other processing entities should be
started. Another processing entity is called a \EM{contingency}.
Rollforward recovery is captured through the irreflexive function
$Rollf\colon \NonSyncs \Func \set{\mbox{`\SF{Redo}'}} \times \NonSyncs$.
Contingencies are defined through the relation
$\Cont = \NonSyncs \times \NonSyncs \times \N \intersect \Rollf$. 
This denotes the fact that a contingency for 
a processing entity $x$ is another processing entity $y$, i.e.\ $x \neq y$, if
$x$ has failed to start $n$ times. Depending on the number, a number of
contingencies are possible. If n equals infinity, then then $x$ is said to be
\EM{forcible}, i.e.\ it has to start at some stage. Forcibility is a requirement
in distributed transaction processing, particularly for compensations (see
below).

The recovery strategy for failures involving non-failure generated aborts
is a \EM{rollback recovery} whereby an either an \EM{undo} or a 
\EM{compensation} or in fact nothing is applied to each processing entity
within the execution path of that service state. An undo simply removes
object store updates provided that they are uncommitted, i.e. the process
was not committed on completion of its execution. A compensation is
required if a commit occurred. Nothing occurs if the effect of the process
is considered uncritical. Rollback is defined through
the irreflexive function 
$\Rollf\colon \NonSyncs \Func \set{\mbox{`\SF{Undo}'},\mbox{`\SF{Null}'}} 
\times \NonSyncs \times \NonSyncs$. Compensations 
are defined through the relation $\Comp = \NonSyncs \times \NonSyncs \intersect
\Rollf$. 

\subsection{Service Model}
\label{sm-tech}

A service model allows the descriptive aspects of a business transaction
workflow to be modelled. Said otherwise, services encapsulate
business transaction workflows. Services are modelled as objects
where object behaviour modelling enables a declaritive approach for
service specification. 

Services have a set $\States$ of states which define its lifecycle,
e.g.\ \SF{Lodged}, \SF{Initial review passed}.
Of course, states are named, given by the function $\SName\colon
\States \Func \SNames$. States are used to determine what processing 
entities of the workflow are triggered for given events. As a result 
of that part of the workflow being executed, a transition between 
the states occurs. In general, transitions are defined through the
relation $\Transition \subseteq \Powerset^+(\States x \States)$. The
powerset indicates that multiple transitions can exist between the
same set of states (recall states \SF{(4)} and \SF{(5)}.

Attached to each transition is an ECA rule. This represents an event
specification.

\section{Conclusion}
\label{conc}

The Application Database being absorbed into the Tenures Administration
System

\subsection*{Acknowledgements}

The authors would like to thank the following staff from the Department
of Natural Resources: Dennis Kelly for information about on road closures, to
Ray Richardson for information about tenure administration and to
Nev Cumerford for information about the department's business 
services, activities and databases.

\begin{scriptsize}
   \BIBLIOGRAPHY{alpha}
\end{scriptsize}
\newpage
\tableofcontents
\end{document}